\documentclass[aps,prl,twocolumn,showpacs,groupedaddress]{revtex4}
\usepackage{graphicx,color}
\begin{document}
\title{Minimal energy packings and collapse of sticky tangent hard-sphere polymers} 
\author{Robert S. Hoy}
\author{Corey S. O'Hern}
\affiliation{Department of Mechanical Engineering, Yale University, New Haven, CT 06520-8286} 
\affiliation{Department of Physics, Yale University, New Haven, CT 06520-8120} 
\pacs{61.46.Bc,64.70.km,82.70.Dd,64.60.Cn} 
\date{\today}

\begin{abstract}
We enumerate all minimal energy packings (MEPs) for small single
linear and ring polymers composed of spherical monomers with contact
attractions and hard-core repulsions, and compare them to 
corresponding results for monomer packings.  We define and identify
``dividing surfaces" in polymer packings, which reduce the number of
arrangements that satisfy hard-sphere and covalent bond constraints.
Compared to monomer MEPs, polymer MEPs favor intermediate
structural symmetry over high and low symmetries.  We also examine the
packing-preparation dependence for longer single chains using
molecular dynamics simulations.  For slow temperature quenches, chains
form crystallites with close-packed cores.  As quench rate
increases, the core size decreases and the exterior becomes more
disordered.  By examining the contact number, we connect 
suppression of crystallization to the onset of isostaticity in disordered
packings.  These studies represent a significant step forward in our
ability to predict how the structural and mechanical properties of
compact polymers depend on collapse dynamics.
\end{abstract}
\maketitle

Over the past several decades significant research
activity has focused on understanding dense packings of hard spheres,
since they serve as model systems for atomic and colloidal liquids and
glasses, jammed granular media, and compressed foams and
emulsions.  An intriguing property of hard-sphere systems is that they
can be prepared in crystalline, partially ordered, and amorphous
packings~\cite{longj}.  Packings of `sticky' hard spheres with contact
attractions have been used to investigate self-assembly of
colloidal particles with depletion attractions.  Arkus \textit{et al.}
recently combined graph theory and geometrical techniques
\cite{arkus09,arkusthesis} to enumerate minimal energy packings
(MEPs), \textit{i.e.} those with the maximum number of contacts, for
$N \leq 10$ sticky hard spheres.  Their predictions agreed with
experiments on attractive colloids~\cite{arkus10}.

However, there have been few studies of packings of sticky tangent
hard-sphere {\it polymers}, which can model polymer collapse, protein
folding, and protein interactions \cite{pham}.  Recent
simulations \cite{laso09,lopatina10} and experiments \cite{zou09} have
investigated polymer packings; however, they considered non-sticky
spheres with only hard-core repulsions, where free volume, not 
energy, is relevant.  Thus, there is little understanding of how
covalent bond and chain uncrossability constraints affect structural
and mechanical properties of sticky hard-sphere polymer packings and
the probabilities with which these occur.

In this Letter, we perform exact enumeration studies of MEPs for
sticky, tangent, monodisperse hard-sphere polymers (both linear and
cyclic) and contrast the results with those for sticky hard spheres
without polymer constraints.  Our studies begin to address several
overarching questions: 1) How do the probabilities for obtaining
polymer MEPs differ from those for sticky hard-sphere MEPs? and 2) How
do the properties of single compact polymers depend on collapse
dynamics, {\it e.g.} do they collapse into crystalline or amorphous
clusters?

Our results show that polymer constraints reduce the ways in
which hard spheres can be arranged into MEPs, and the strength of
this effect varies for different macrostates ({\it i.e.} structurally
distinct packings).  We demonstrate that the large reduction in
the number of arrangements may be understood in terms of
\textit{dividing surfaces}.  These split polymer packings into
disjoint regions and eliminate particle-label permutations that do not
correspond to polymer chains.  We find that polymer MEPs with
intermediate structural symmetry are more frequent relative to the
monomer case, where entropy favors low symmetry
MEPs~\cite{arkus10}.

In addition, using molecular dynamics (MD) simulations of temperature
quenches at various rates $\dot{T}$, we show that single chains
display glassy dynamics during collapse, and that the final polymer
packings depend on $\dot{T}$.  In the slow quench rate limit, the
chains undergo a sharp \cite{foot7} transition to crystallites, with a jump in the
energy and number of contacts $N_c$ (including covalent bonds) at
temperature $T=T_{\rm melt}$.  The crystallites possess a
close-packed core surrounded by a ``surface'' whose size and disorder
increase with $|\dot{T}|$. For slow quenches, $N_c$ at $T_{\rm melt}$
jumps from below the minimal number $N_c^{\rm min}=3N-6$ required for
mechanical stability~\cite{donev05} to $N_c^{\rm slow}$, where a
significant fraction of the monomers possess $12$ contacts.  In the
large $|\dot{T}|$ limit, the clusters are disordered with 
$\lesssim N_c^{\rm min}$ contacts even as $T\to0$, showing that
rigidification can hinder crystallization.

\begin{table*}[htbp]
\caption{Statistics for MEPs with $N$ spheres and $N_c$ contacts.  $M$
is the number of macrostates, $f_r$, $f_p$, and $f_m$ are the fraction
of microstates obeying minimal rigidity constraints that also satisfy
hard-sphere constraints, respectively for rings, linear polymers, and
monomers, and $\Omega_{r}$, $\Omega_{p}$, and $\Omega_{m}$ are the
total numbers of microstates satisfying both minimal rigidity and
hard-sphere constraints.  Values for $f$ and $\Omega$ do not account for chiral twins \cite{arkus09}.  In agreement with \cite{arkus09}, we find $1$ and $4$ floppy macrostates (in
the $k\rightarrow \infty$ limit~\cite{epaps}), respectively for $N=9$ and ($N=10$, $N_c = 24$).  However, we find $2$ and $55$ more rigid macrostates$^{*,\#}$ for these cases \cite{foot6,epaps}.
Adjacency matrices and coordinate solutions for all microstates are available online
\cite{onlinestuff}. $-$ indicates data not available.}
\begin{ruledtabular}
\begin{tabular}{lcccccccccc}
$N$ & $N_c$ & $M$ & $f_{r}$ & $f_{p}$ & $f_{m}$ & $f_{r}/f_{m}$ & $f_{p}/f_{m}$ & $\Omega_{r}$ & $\Omega_{p}$ & $\Omega_{m}$\\
5 & 9 & 1 & 1 & 1 & 1 & 1 & 1 & 5 & 6 & 10\\
6 & 12 & 2 & 0.435 & 0.463 & 0.494 & 0.88 & 0.94 & 34 & 50 & 195\\
7 & 15 & 5 & 0.102 & 0.114 & 0.134 & 0.76 & 0.85 & 273 & 486 & 5712\\  
8 & 18 & 13 & $1.66\cdot10^{-2}$ & $1.91\cdot10^{-2}$ & $2.45\cdot10^{-2}$ & 0.68 & 0.78 & 2668 & 5500 & 231840\\
9 & 21 & $52^{*}$ & $1.40\cdot10^{-3}$ & $2.46\cdot10^{-3}$ & $3.34\cdot10^{-3}$ & 0.42 & 0.74 & 30663 & 71350 & 12368160\\
10 & 24 & $278^{*,\#}$ & $2.21\cdot10^{-4}$ & $2.55\cdot10^{-4}$ & $-$ & $-$ & $-$ & 426590 & 1093101  & $-$ \\
10 & 25 & $3$ & $2.05\cdot10^{-6}$ & $1.98\cdot10^{-6}$ & $-$ & $-$ & $-$ & 5905 & 12138 & $-$ 
\end{tabular}
\end{ruledtabular}
\label{tab:omega2}
\end{table*}

We first describe exact enumeration methods for monomer and polymer
MEPs~\cite{epaps}.  To generate possible packings for a given number
of spheres $N$ and contact number $N_c$, we iterate over all $N\times N$ adjacency
matrices $\bar{A}$ satisfying $\sum_{j>i} A_{ij} = N_c$.
The elements of $\bar{A}$ are $1$ for contacting
particles, and $0$ for non-contacting particles and 
diagonal entries.
Covalent bonds link sticky spheres to form a polymer chain with length $N$; $A_{i,i+1}=1$ for $1 \leq i < N$ for linear chains, and additionally $A_{1,N}=1$ for rings.  
The distinction between permanent covalent and thermally fluctuating 
noncovalent bonds is not important for static packings; we include both types in $N_c$.

We enumerate all adjacency matrices satisfying the above conditions
and then identify those that also fulfill hard-sphere and minimal rigidity
constraints.  Hard-sphere constraints imply that the
center-to-center distances $r_{ij}$ between unit spheres $i$ and $j$
obey $r_{ij} \geq 1$, where the equality holds for contacting pairs.
Necessary conditions for rigidity are that each monomer possesses at
least three contacts and $N_c \geq N_c^{\rm min}$~\cite{jacobs95}.

To enforce these constraints, we implemented geometrical rules
developed by Arkus {\it et al.}~\cite{arkus09,arkusthesis} that
eliminate invalid adjacency matrices.  For the remaining
configurations, we solved the system of quadratic equations
\begin{equation}
|\vec{r}_{i}-\vec{r}_{j}|^2 = d_{ij}^2
\label{eq:variety}
\end{equation} 
for sphere positions ${\vec r}_i$ to an accuracy of $10^{-9}$.  We
also calculated the dynamical matrix (all second derivatives of the
energy in Eq.\ (\ref{eq:perturbed}) with respect to monomer displacements)
for all configurations, which allowed us to identify rigid (with
$3N-6$ nonzero eigenvalues) and floppy configurations~\cite{jacobs95}
(with fewer nonzero eigenvalues) \cite{epaps}.

From this procedure, we obtain microstates and macrostates for a given
$N$ and $N_c$ that satisfy hard-sphere and minimal rigidity constraints and
the relevant polymeric constraints.
Each macrostate is characterized by an adjacency matrix that is
nonisomorphic to and a set of interparticle distances $\{r_{ij}\}$
that is different from those characterizing other macrostates \cite{epaps}.  With
this definition, no macrostate can be rotated or reflected such that
it yields a different macrostate.  Every connected sticky hard-sphere
macrostate admits a linear polymer macrostate~\cite{biedl01}.  Thus,
sticky-sphere and linear polymer packings have identical macrostates.  We
have also verified this for ring packings for $N \leq 10$.

A microstate is a particular labeling of the
particles $1$ through $N$ that comprise a $N$-particle
macrostate with $N_c$ contacts.  
Many microstates correspond
to each macrostate due to particle permutations for monomer packings
\cite{arkus10}, and for polymers, the multiple possible paths through a
given macrostate.  The total number of microstates $\Omega_m$,
$\Omega_p$, and $\Omega_r$ is given by the sum of microstates
for each macrostate for monomers, linear polymers,
and rings, respectively \cite{epaps}.  For monomer packings, which lack covalent bonds, the number of microstates
for each macrostate (ignoring chirality) is given simply by a
geometric factor $\Omega_m^{i} = P_i$, where $P_i$ is the number of
allowed permutations of particle indices for macrostate $i$ \cite{arkusthesis}. For
polymer packings, the number of microstates is not given by this
relation since one must ensure that particle indices are consecutive.

\begin{figure}
\vspace{-14pt}
\includegraphics[width=3.0in]{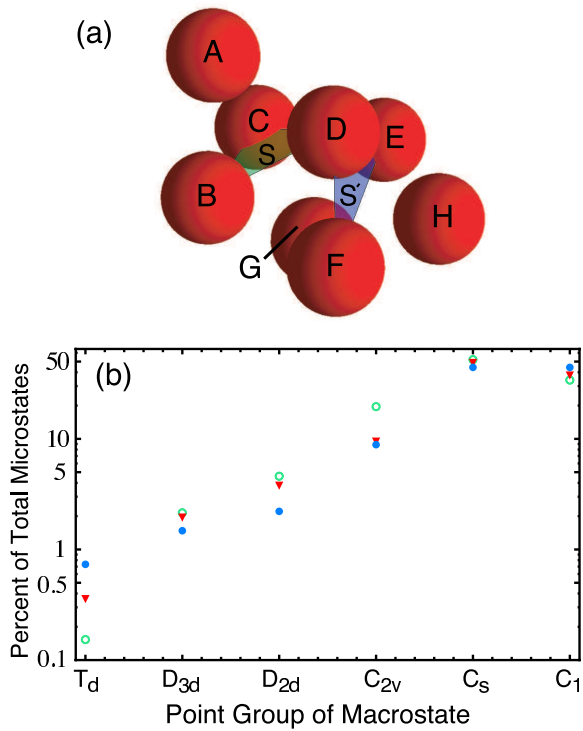}
\caption{(a) Schematic of dividing surfaces $S$ and $S'$ (colored
triangles formed by monomers ($B$, $C$, $D$) and ($D$, $E$, $F$),
respectively) in a macrostate for $N=8$.  For $S$, region $J$ consists
of monomer $A$ and region $K$ of monomers ($E$, $F$, $G$, $H$), or
vice versa.  (b) Fraction of microstates for packings from each
symmetry group for cyclic (open circles) and linear (downward
triangles) polymers, and monomers (filled circles) with $N=8$. Results
in (b) do not account for chiral structures.}
\label{fig:dividing}
\vspace{-0.15in}
\end{figure}

Exact enumeration results are displayed in
Table~\ref{tab:omega2}, which shows the number of macrostates $M$,
fraction $f$ of adjacency matrices with $N_c$ contacts obeying minimal
rigidity that also satisfy hard-sphere constraints, and $\Omega_m$,
$\Omega_p$, and $\Omega_r$ for $5 \leq N \leq 10$.  
$f$ is the probability to obtain a packing
for an `ideal' protocol that samples adjacency matrices uniformly.
From Table~\ref{tab:omega2}, we see that $f$ decreases approximately
exponentially with $N$ for $N \ge 5$, and even faster for $N > 9$.
Part of the reason for the strong decrease in $f$ between $N=9$ and
$10$ is the decrease in macrostates from $52$ to $3$.  This
occurs because $N=10$ MEPs possess $N_c = N_c^{\rm min}+1$, which
exceeds the number of degrees of freedom.  Eq.~\ref{eq:variety} is then
overconstrained, and its solutions possess special
symmetries.  The increase in $N_c$ signals the onset of crystal
nucleation, and the formation of a close-packed core.  The ability to
enumerate the numbers of isostatic ($N_c = N_c^{\rm min}$) and
hyperstatic ($N_c > N_c^{\rm min}$) packings will yield insight into
systems where glass and crystallization transitions compete.

For the $N$ studied here, hard-sphere constraints are more difficult
to satisfy for minimally rigid polymer packings compared to monomer
packings: $f_{r} < f_{p} < f_{m}$~\cite{foot5}.  A key mechanism for
the reduction in $f$ is the occurrence of ``dividing surfaces'' in
polymer packings.  A dividing surface is any minimal subset of a
connected cluster of contacting monomers that geometrically splits it
into two.  Any polymer path that traverses a dividing surface that
does not also topologically divide the polymer is blocked and
invalid. Specifically, if $m$ consecutive monomers $i+1,\ldots,i+m$
occupy an $m$-monomer dividing surface $S$, any polymer path where the sets of
monomers $J$ and $K$ divided by $S$ are anything other than
$1,2,\ldots,i$ and $i+m+1,i+m+2,\ldots,N$ (or vice versa) is
blocked. In other words, any path that starts in $J$, enters $S$, and
traverses it (passes through all monomers in $S$) is blocked unless it
traverses all monomers in $J$ before entering $S$.  
Fig.\ \ref{fig:dividing}(a) schematically depicts the sets $J$ and $K$ and two dividing surfaces for
a $N=8$ macrostate. 
By definition, blocking does not occur in monomer packings.

In Table~\ref{tab:omega2}, we see that the blocking effect increases
sharply with $N$ since $f_r/f_m$ and $f_p/f_m$ decrease significantly.
Blocking also reduces \cite{foot5} the fractions of allowed ring
microstates relative to those for linear polymers $f_{r}/f_{p}$
since rings do not possess chain ends.  Another clear feature in
Fig.~\ref{fig:dividing}(b) is that blocking changes the relative
frequencies with which macrostates of different symmetries are
populated.  Ring and linear polymer packings are more likely to possess
intermediate symmetry than monomer packings, whereas the opposite is
true for macrostates with the lowest and highest symmetries.  Highly
symmetric macrostates possess many distinct blocking surfaces, and
low symmetry macrostates possess a surplus of closed trimers as
shown in Fig.~\ref{fig:dividing}(a).

The enumeration studies illustrate an interesting
competition between energy and entropy in large systems.  For $N\geq
10$, MEPs are overconstrained with $N_c > N_c^{\rm min}$.
This suggests that if the system becomes trapped in a metastable state
({\it e.g.} with $N_c = N_c^{\rm min}$), rearrangements into MEPs will
be slow because of their low entropy.  Thus, glassy dynamics in single
polymer chains should be observable in systems quenched at varying
rates.  For $k_B T \gg |\epsilon|$, where $-\epsilon$ is the contact
energy, polymers will adopt random-coil configurations with $N_c \ll
N_c^{\rm min}$.  As the polymer is cooled, one expects quench rate
effects to become important when $N_c \simeq N_c^{\rm
min}$~\cite{huerta02b}.

To demonstrate glassy dynamics for single linear polymer chains, we employ  
MD simulations in which monomers interact via the potential energy
\begin{equation} 
U_{\rm harm}(r) = \bigg{\{}\begin{array}{ccc}
-\epsilon + \frac{k}{2}(\frac{r}{D}-1)^{2} & , & r < r_c\\
0 & , & r > r_c
\end{array},
\label{eq:perturbed}
\end{equation}
where $k$ is the spring constant and $D=1$ is the monomer
diameter. The temperature $T$ is controlled via a Langevin thermostat.
The unit of time is $\tau=\sqrt{mD^2/\epsilon}$, where $m$ is the
monomer mass.  The cutoff radius $r_c/D = \infty$ for covalently
bonded monomers and $1 + \sqrt{2\epsilon/k}$ for noncovalently bonded
monomers.  $U_{\rm harm}$ reduces to the energy for tangent sticky
hard spheres~\cite{sticky} in the limit $k\to\infty$ and possesses the
same MEPs.  For $N \leq 10$, the MEPs from simulations agree with
those from complete enumeration for $k \gtrsim 1600\epsilon$ ($r_c
\lesssim 1.04D$).

\begin{figure}
\includegraphics[width=3.0in]{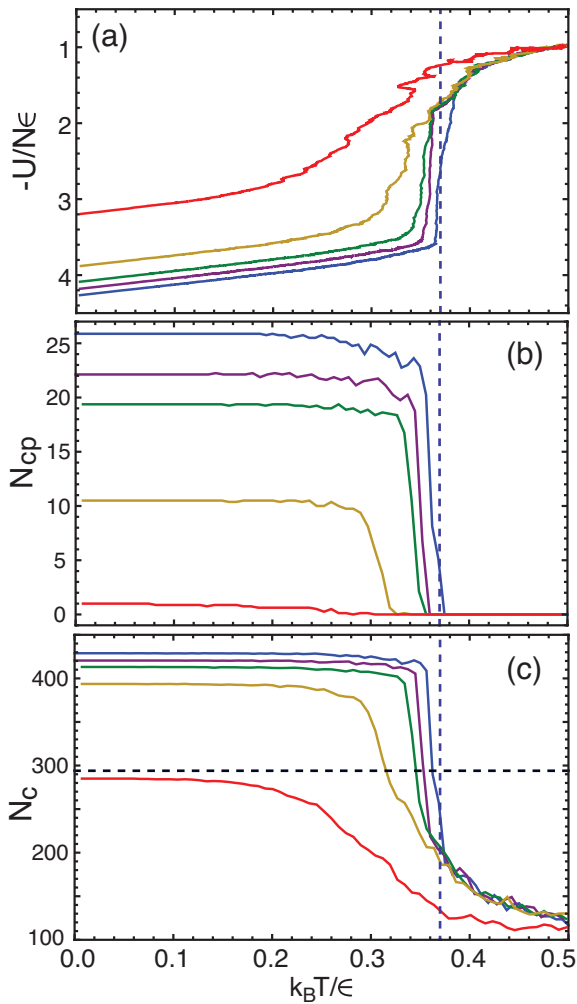}
\vspace{-0.1in}
\caption{(a) Potential energy per particle ($-U/N\epsilon$) (b) number of
particles with $12$ contacts ($N_{cp}$), and (c) total number of
contacts ($N_c$) versus $k_B T/\epsilon$ for single linear polymers
with $N=100$ at different quench rates.  Data (top to bottom,
panels b-c; bottom to top, panel a) are for quench rates $k_B\dot{T}
\tau/\epsilon = -10^{-3}$, $-10^{-4}$, $-10^{-5}$, $-10^{-6}$ and
$-10^{-7}$.  The critical quench rates are $|k_B \dot{T}^{*}/\epsilon|
\sim 10^{-7}/\tau$ and $|k_B \dot{T}^{**}/\epsilon| \sim
10^{-3}/\tau$.  All results are averaged over several independent
initial configurations.  The horizontal (vertical) dotted lines
indicate $N_c = N_c^{\rm min}$ ($k_B T/\epsilon = 0.37$).}
\label{fig:crystglasscp}
\vspace{-0.15in}
\end{figure}

Figure \ref{fig:crystglasscp}(a) shows the potential energy per
particle $-U/N\epsilon$ for different quench rates $\dot{T}$.  At low
$\dot{T}$, a sharp transition between coils and crystallites \cite{taylor09,foot7} is
observed at $T_{\rm melt} \simeq 0.37\epsilon/k_B$.  The crystallites consist of a close-packed
core with $N_{cp}$ monomers (each with 12 contacts) and a less-ordered
exterior. The crystallization transition corresponds (Fig.\ \ref{fig:crystglasscp}(b)) to a sharp
transition in $N_{cp}$, which implies a change of symmetry within the
core, from liquid-like to close-packed.

At higher rates, the dynamics becomes glassy near $T_{\rm melt}$, and
the systems do not approach the ground state energy even as $T\to
0$. We associate the suppression of crystallization with the onset of
rigidity. Evidence for this is given in Fig.\
\ref{fig:crystglasscp}(c).  The data show two ``critical'' quench
rates: $\dot{T}^{*}$ and $\dot{T}^{**}$.  For $|\dot{T}| <
|\dot{T}^*|$, the jump in $N_c$ and $N_{cp}$ resembles a first-order
transition.  For $|\dot{T}| > |\dot{T}^{**}|$, the systems do not form
minimally rigid clusters even at $T=0$.  Even though the critical
rates and $T_{\rm melt}$ are $N$-dependent, the trends are clear.  For
$N=100$ systems, we estimate $|k_B \dot{T}^{*}/\epsilon| \sim
10^{-7}/\tau$ and $|k_B \dot{T}^{**}/\epsilon| \sim 10^{-3}/\tau$.

\begin{figure}[htbp]
\vspace{-16pt}
\includegraphics[width=3.1in]{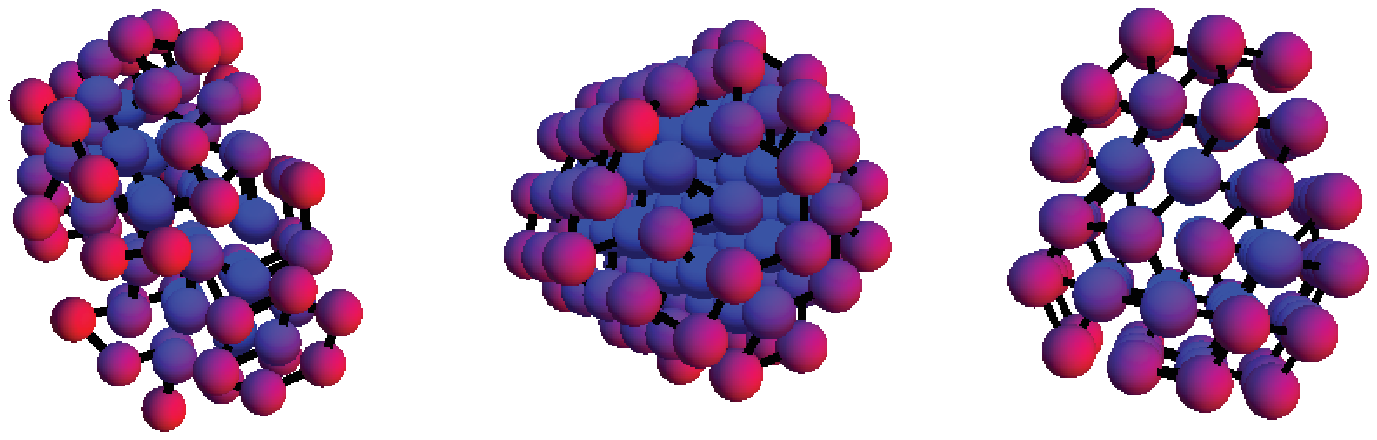}
\vspace{-0.26in}
\caption{Collapsed structures at $T=0$ for a single $N=100$ linear polymer
using two quench rates: $k_B \dot{T} \tau/\epsilon =-10^{-4}$
(left) and $-10^{-7}$ (middle, right).  The packing in the right panel
is rotated compared to that in the middle panel to show its hexagonal
planes.}
\label{fig:quenchedhundred}
\vspace{-0.15in}
\end{figure}

The effects of quench rate on end states of quenches to $T=0$ are
visualized in Fig.\ \ref{fig:quenchedhundred}.  Monomers are
color-coded by the number of contacts; dark blue (red) indicates close
packing ($\ll 12$ contacts). The left panel shows a typical
configuration after a fast quench with $k_B
\dot{T}=-10^{-4}\epsilon/\tau$; we see a small close-packed core
surrounded by a disordered exterior.  The middle and right panels show
a collapsed structure at $T=0$ from a slow quench
($k_B\dot{T}=-10^{-7}\epsilon/\tau$).  The close-packed core is much
larger, and the exterior is more crystalline.  The large gaps visible
in the rightmost panel indicate the order is hcp, and the structure is
stack-faulted~\cite{karayiannis09prl}.

We examined minimal energy packings of sticky tangent hard-sphere
linear and cyclic polymers, and compared them to monomer packings for
small $N$. The packings are the same, but polymer packings possess
significantly smaller entropies compared to monomer packings due to
dividing surfaces, which arise from covalent-bond constraints.
Entropic suppression via blocking is strongest for structures of both
very high and low symmetry.  In both monomer and polymer
cases, the fraction of states satisfying hard-sphere constraints
decreases at least exponentially with increasing $N$, and faster when
$N_c > N_c^{\rm min}$.  We also performed MD simulations of single
linear chains with larger $N$, which link glassy dynamics to the onset
of rigidity.  This work sets the stage for future studies that
investigate whether cooperative dynamics from chain connectivity
and uncrossability constraints improves or impedes glass-forming
ability of single polymers compared to colloidal systems.

We thank V.\ N.\ Manoharan for helpful discussions.  Our results were
obtained using the Boost Graph Library, a modified version of N.\
Arkus' structure solver~\cite{arkusthesis}, and LAMMPS
\cite{plimpton95}.  Support from NSF Award No.\ DMR-0835742 and an
Anderson Fellowship from Yale University is gratefully acknowledged.

\section{Supplementary Material for ``Minimal energy packings and collapse of sticky tangent hard-sphere polymers''}

In this supplementary material, we provide additional details
concerning the methods employed to: (1) enumerate exactly all micro- and
macrostates and (2) assess the rigidity for monomer and polymer
packings composed of sticky, monodisperse tangent hard spheres.

\subsection{Exact enumeration method}
\label{exact}

The exact enumeration method consists of several steps including
looping over all adjacency matrices satisfying the appropriate
constraints for monomer and polymer packings, identifying those
adjacency matrices that satisfy hard-sphere constraints, and then
solving for their Euclidean positions.
  
The adjacency matrix $\bar{A}$ for a $N$-particle system is a $N\times
N$ symmetric matrix whose elements are $1$ for contacting particles and
$0$ for noncontacting particles.  For monodisperse hard spheres,
\begin{equation}
\begin{array}{lllll}
A_{ij} & =  & 1 & \textrm{if} & r_{ij} = D,\\
A_{ij} & =  & 0 & \textrm{if} & r_{ij} > D,
\end{array}
\label{eq:adjmatone}
\end{equation}
where $r_{ij} = |\vec{r}_{i} - \vec{r}_{j}|$ is the center-to-center
distance between spheres $i$ and $j$ and $D$ is their diameter.  By
convention, the diagonal entries satisfy $A_{ii} = 0$.

The number of permutations for symmetric $N\times N$ matrices
with $N_{const}$ constrained elements ({\it e.g.} covalent bonds) is
\begin{equation}
\label{constraint}
P_{\rm const} = \displaystyle\frac{[(N^{2} - N)/2]!}{N_{\rm const}![(N^{2} -
N)/2-N_{\rm const}]!},
\label{eq:constrainedpermutations}
\end{equation}
where the $(N^2-N)/2$ terms arise because we only need to consider entries
above the diagonal since $\bar{A}$ is symmetric.  The number of
permutations $P_{\rm cont}$ for $\bar{A}$ with $N_c$ sticky-sphere
contacts has the same form as Eq.~\ref{constraint}:
\begin{equation}
P_{\rm cont} = \displaystyle\frac{[(N^{2} - N)/2]!}{N_{c}!((N^{2} - N
)/2 - N_{c})!}.
\label{eq:binarypermcoll}
\end{equation}

For the packings considered in this study, the constrained elements
correspond to covalent bonds, which are fixed to be $1$ not $0$.  Thus, the
number of permutations is
\begin{equation}
\begin{array}{llc}
P_{\rm cct}(N_{\rm const}) & = & 
\frac{\left[\frac{N^{2} - N}{2} - N_{\rm
const}\right]!}{(N_{c}-N_{\rm const})! \left[\frac{N^{2} - N}{2} -
(N_{c}-N_{\rm const})\right]!}, 
\end{array}
\label{eq:constrainedcontacts}
\end{equation}
where $N_{\rm const} = 0$, $N-1$, and $N$, for monomers, linear
polymers, and rings, respectively.  Specifically, for linear and ring
polymers, $A_{i,i+1}=1$ for $1 \leq i < N$.  Rings must also satisfy
the constraint $A_{1,N}=1$.  In contrast, monomer packings
do not have explicitly constrained off-diagonal elements.

Thus, the numbers of adjacency matrices for monomers, linear polymers,
and rings \cite{foot1} for $N$ particles and $N_c$ contacts (including
covalent bonds for polymers) are $P_m = P_{\rm cont}$,
\begin{equation}
\begin{array}{rl}
P_{lp} = & P_{\rm cct}(N-1), \\
 = & \displaystyle\frac{[\frac{N^{2} - 3N +
  2}{2}]!}{(N_{c}-(N-1))!(\frac{N^{2} - 3N+2}{2} - (N_{c}-(N-1))!}, \nonumber
\end{array}
\label{eq:binarypermpol}
\end{equation}
and
\begin{equation}
\begin{array}{rl}
P_{r} & = P_{\rm cct}(N), \\ 
= & \displaystyle\frac{[(\frac{N^{2} -
3N}{2}]!}{(N_{c}-N)!(\frac{N^{2} - 3N}{2} - (N_{c}-N))!}. 
\end{array}
\label{eq:binarypermring}
\end{equation}

We then loop through all
permutations $P_m$, $P_{lp}$, and $P_r$ of $\bar{A}$ for a given $N$
and $N_{c}$.  Since the entries of $\bar{A}$ are ones and zeros,
different adjacency matrices correspond to unique binary numbers ({\it
i.e.} each microstate corresponds to a particular adjacency matrix
and unique binary number).  We enumerate all binary numbers 
using sequential binary permutations (from the C++
Standard Template Library \textit{next\_permutation()} function).
Note that considering polymers leads to an exponential reduction in
the effort required for exact enumeration.

Since we are interested in minimal energy packings (MEPs)---those with
the maximum number of contacts $N_c$, we focus on packings with $N_c
\ge N_c^{\rm min}$, where $N_c^{\rm min} \equiv 3N-6$ is the minimal
number of contacts required for rigidity.  For $4 \leq N \leq 9$, we
verified that no microstates with $N_c > N_c^{min}$ satisfy hard
sphere constraints (as shown previously~\cite{arkus09}), and for
$N = 10$ no microstates with $N_c = N_c^{min} + 2$ exist, {\it i.e.} 
MEPs for $4 \leq N \leq 9$ possess $N_c = N_c^{\rm min}$ and for 
$N = 10$ possess $N_c = N_c^{\rm min}+1$.

To eliminate adjacency matrices that do not satisfy hard-sphere
constraints, we employed the complete set of geometrical rules for the
adjacency matrices for $N\leq7$ provided in Ref.~\cite{arkusthesis}.
Specifically, we implemented rules $1$-$12$, $14-18$ outlined on pages 295-318.
However, we did not employ the triangular bipyramid rule (discussed on
pages 40-48 of \cite{arkusthesis}), neither the version for iterative
packings nor that applied to new seeds.  Instead, for $N>7$ we
inserted those minimally rigid adjacency matrices not rejected by the
geometrical rules into a modified version of Arkus' Euclidean
structure solver.  The structure solver makes a random initial guess for
particle coordinates and then uses Newton's method to solve the
contact equations implied by the adjacency matrix, while enforcing
hard sphere constraints.  We checked for convergence of the
structure solver by increasing the maximum number of `initial guesses'
for the coordinates; $3N^3$ initial guesses are sufficient to solve
all structures to an accuracy in positions of $10^{-6}$.

Closely associated with $\bar{A}$ is the distance matrix $\bar{D}$
whose elements are $D_{ij} = r_{ij}$.  For $N < 10$ and $N=10$, $N_c =
25$, nonisomorphic $\bar{A}$ (identified using the Boost Graph
Library's \textit{isomorphism()} function) correspond to different
macrostates because the elements of $\bar{D}$ are different
\cite{foot2}.  However, for $N=10$, $N_c=24$ graph (non)isomorphism
(as given by \textit{isomorphism()}) is insufficient to completely
distinguish macrostates.  While there are $286$ nonisomorphic graphs,
$8$ of these produce coordinate solutions that are identical to those
produced by other graphs.  All eight of these correspond to
`switching' a noncovalent bond in such a way that the same coordinate
solution is produced.  This reduces the total number of macrostates to
$278$.

For example, graphs $128$ and $158$ produce the same set of
coordinates \cite{onlinestuff}, but differ in that the former has a
noncovalent bond between particles $1$ and $6$, while the latter
possesses a noncovalent bond between particles $4$ and $7$.  We assume
that graphs $128$ and $158$ correspond to the same macrostate because
they possess the same $\bar{D}$, but (since the particles are
distinguishable) the coordinate solutions correspond to different
microstates \cite{foot3}.  We therefore have retained all microstates
for all $286$ graphs, but assign them to $278$ macrostates.  Adjacency
matrices for graphs $128$ and $158$ and their common coordinate solution 
are shown below in (\ref{eq:graphcoords128158}); red entries indicate the `switched' bond.

\begin{equation}
\begin{array}{lc}
\textrm{Graph 128:} &
\left[\begin{array}{cccccccccc}
0 & 1 & 1 & 1 & 0 & \color{red}{1} & 1 & 1 & 1 & 1\\
1 & 0 & 1 & 0 & 1 & 1 & 0 & 0 & 0 & 1\\
1 & 1 & 0 & 1 & 1 & 0 & 0 & 0 & 1 & 1\\
1 & 0 & 1 & 0 & 1 & 1 & 0 & 1 & 0 & 0\\
0 & 1 & 1 & 1 & 0 & 1 & 0 & 0 & 0 & 0\\
\color{red}{1} & 1 & 0 & 1 & 1 & 0 & 1 & 0 & 0 & 0\\
1 & 0 & 0 & 0 & 0 & 1 & 0 & 1 & 0 & 0\\
1 & 0 & 0 & 1 & 0 & 0 & 1 & 0 & 1 & 0\\
1 & 0 & 1 & 0 & 0 & 0 & 0 & 1 & 0 & 1\\
1 & 1 & 1 & 0 & 0 & 0 & 0 & 0 & 1 & 0
\end{array}\right]\\
& \\
\textrm{Graph 158:} &
\left[\begin{array}{cccccccccc}
0 & 1 & 1 & 1 & 0 & 0 & 1 & 1 & 1 & 1\\
1 & 0 & 1 & 0 & 1 & 1 & 0 & 0 & 0 & 1\\
1 & 1 & 0 & 1 & 1 & 0 & 0 & 0 & 1 & 1\\
1 & 0 & 1 & 0 & 1 & 1 & \color{red}{1} & 1 & 0 & 0\\
0 & 1 & 1 & 1 & 0 & 1 & 0 & 0 & 0 & 0\\
0 & 1 & 0 & 1 & 1 & 0 & 1 & 0 & 0 & 0\\
1 & 0 & 0 & \color{red}{1} & 0 & 1 & 0 & 1 & 0 & 0\\
1 & 0 & 0 & 1 & 0 & 0 & 1 & 0 & 1 & 0\\
1 & 0 & 1 & 0 & 0 & 0 & 0 & 1 & 0 & 1\\
1 & 1 & 1 & 0 & 0 & 0 & 0 & 0 & 1 & 0
\end{array}\right]\\
& \\
\textrm{Coordinates:} & \left[
\begin{array}{ccc}
 0 & 0 & 0 \\
 0 & 1 & 0 \\
 -0.866025 & 0.5 & 0 \\
 -0.57735 & 0 & -0.816497 \\
 -0.57735 & 1 & -0.816497 \\
 0.288675 & 0.5 & -0.816497 \\
 0.288675 & -0.5 & -0.816497 \\
 -0.481125 & -0.833333 & -0.272166 \\
 -0.7698 & -0.333333 & 0.544331 \\
 -0.288675 & 0.5 & 0.816497
\end{array}
\right]\\
& \\
\end{array}
\label{eq:graphcoords128158}
\end{equation}

\subsection{Rigidity analysis}

The rigidity of all macrostates generated by our exact enumeration
algorithm was assessed by calculating the eigenvalues of the dynamical
matrix~\cite{gao}, assuming the following harmonic interparticle
potential
\begin{equation} 
U_{\rm harm}(r) = 
-\epsilon + \frac{k}{2}\left(\frac{r}{D}-1\right)^{2},
\label{eq:perturbed2}
\end{equation}
where $D=1$ and $k$ must be large enough so that no `2nd-nearest
neighbors' interact.  We have examined the number of rigid macrostates
as a function of $k$.  For sufficiently high precision in the
coordinates (one part in $10^{6}$ or better), all macrostates for
$4 \leq N \leq 8$ are rigid ({\it i.e.} the dynamical matrix possesses
$3N-6$ positive eigenvalues) for $k/\epsilon > 10$.  For the eigenvalue
threshold, we assumed that eigenvalues $e_i > 10^{-1}$ were nonzero and
positive. For $N = 9$, we find one `floppy' macrostate for all
$k/\epsilon > 400$, consistent with Ref.\ \cite{arkus09}.  We also
verified that all $N = 10, N_c = 25$ macrostates possess are rigid.

\begin{figure}
\includegraphics[width=3.0in]{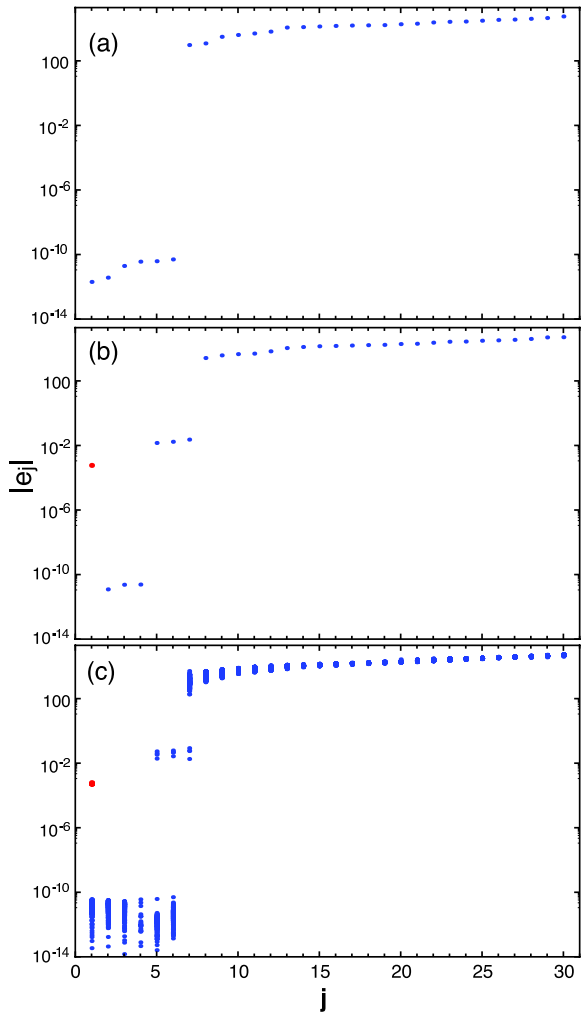}
\caption{Eigenvalue spectra from the dynamical matrix for $N=10$,
$N_c=24$ macrostates.  The eigenvalues $e_j$ are sorted from smallest
to largest, $j=1$ to $3N$.  Eigenvalue spectra are shown for a rigid
and floppy macrostate in (a) and (b) and all $278$ macrostates in (c).
In (a), there are $3N-6$ nonzero eigenvalues and $6$ eigenvalues near
zero that correspond to rigid translations and rotations.  In (b),
there is an extra `zero' eigenvalue (red point) that corresponds to
the floppy mode.}
\label{fig:espectrum}
\end{figure}

For $N = 10, N_c = 24$, however, the results show nontrivial
dependence on the numerical precision of the coordinate solutions.
When we solve for the coordinates to a precision of one part in
$10^{9}$, we find $4$ nonrigid (floppy) macrostates (in agreement with
Ref.\ \cite{arkus09}) for all $k/\epsilon > 400$.  However, when the
coordinates are solved to a precision of only one part in $10^{6}$,
one of the floppy configurations becomes `rigid' in the same range of
$k$ due to insufficient precision.

Figure \ref{fig:espectrum} shows the eigenvalue spectra for the
$N=10$, $N_c= 24$ macrostates for a precision of one part in $10^{9}$
in the coordinate solutions.  The eigenvalues are displayed from
smallest to largest: $j=1$ to $3N$.  The $3N-6$ positive eigenvalues
for rigid macrostates are well-separated from the $6$ eigenvalues that
correspond to rigid translations and rotations as shown in panel (a).
The eigenvalues corresponding to rigid translations and rotations
(indexes $1$-$6$) are zero to within our numerical precision.  For
floppy macrostates, the floppy eigenvalues are also orders of
magnitude below those that correspond to finite-energy normal modes.

Since nonrigid macrostates have floppy modes that can be activated
with no energy cost, each nonrigid macrostate possesses a continuum of
coordinate solutions. However, we have verified that none of the four
floppy macrostates for $N=10$, $N_c = 24$ can be transformed continuously
into one another (without increasing $U_{\rm harm}$).  Specifically,
we have shown that different coordinate solutions corresponding to the
same floppy macrostate differ by $\lesssim 10^{-3} D$, while
transitions between floppy macrostates would require displacements
$\sim D$.


\begin{thebibliography}{25}
\bibitem[{foo({\natexlab{a}})}]{foot7}
\bibinfo{note}{It cannot be rigorously classified as first order since
$N=100$ chains are far from the thermodynamic limit \cite{taylor09}.}
\bibitem[{foo({\natexlab{b}})}]{foot6}
\bibinfo{note}{For $N=10$, $N_c=24$, there are $279$ macrostates; we list
  the $278$ that cannot be formed by breaking one contact of a $N=10$, $N_c=25$
  state~\cite{arkusthesis}.}
\bibitem[{foo({\natexlab{c}})}]{foot5}
\bibinfo{note}{Except for ($N=10$, $N_c=25$); the slight difference may correspond to suppression of $\bar{A}$'s not satisfying minimal rigidity constraints.}
\bibitem[{foo({\natexlab{a}})}]{foot1}
\bibinfo{note}{Eqs.\ \ref{eq:constrainedpermutations}-\ref{eq:binarypermring}
  break down in general if any of the arguments of the factorial terms become
  nonpositive.}
\bibitem[{foo({\natexlab{b}})}]{foot2}
\bibinfo{note}{Another factor that requires the macrostates to be defined with
  care is dimensionality. The Euclidean coordinate solutions are obtained by
  solving $r_{ij}^2 = |\vec{r}_{i}-\vec{r}_{j}|^2 = d_{ij}^2$.
  Some of the $278$ ($N=10$, $N_c=24$)
  macrostates have the same sets of $\{r_{ij}^2\}$ but different sets of
  $\{r_{ij,x}^2, r_{ij,y}^2, r_{ij,z}^2\}$; this occurs because different sets
  of components $r_{ij,x}^2$, $r_{ij,y}^2$, and $r_{ij,z}^2$ (that do not
  correspond to rigid rotations) can give the same sum $r_{ij}^2 = r_{ij,x}^2 +
  r_{ij,y}^2 + r_{ij,z}^2$. This complication does not arise for $N < 10$, but
  we anticipate it will become increasingly important for enumeration of MEPs
  at higher $N$.}
\bibitem[{foo({\natexlab{c}})}]{foot3}
\bibinfo{note}{To see this, note that if the particles are distinguishable, we
  can assign 'colors' to them. Suppose $1 =$ ``red'', $4 =$ ``green'', $6 =$
  ``blue'', and $7 =$ ''yellow''. Thus 128 contains a red-blue bond while 158
  contains a green-yellow bond, and these are distinguishable at the microstate
  level.}
\end{thebibliography}
\end{document}